# On the Problem of Validity of Anderson and Kondo Models in Physics of Strongly Correlated Electron Systems

N. E. Sluchanko

*Prokhorov General Physics Institute, Russian Academy of Sciences, Moscow, 119991 Russia*

E-mail: nes@lt.gpi.ru

We argue that the Anderson and Kondo models turn out to be irrelevant for the description of some strongly correlated electron systems and suggest the mechanism for the formation of many-body states (heavy fermions) being an alternative to the Kondo one. This mechanism involves the quantum tunneling of a heavy particle between the states in the double-well potential.

PACS: 71.27.+a

**1.** One of the main approaches currently used in the physics of strongly correlated electron systems (SCES) for treating the behavior of both an individual magnetic impurity and a periodic array of magnetic centers in metals is based on the Anderson model [1, 2], as well as on the Schrieffer–Wolff transformation [3] leading to its representation in the form of the Kondo model [4]. In the last decade, however, several researchers put forward the arguments challenging this conventional description. In particular, it was shown in [5] that the *Kondo impurity* model is inappropriate for interpreting the behavior of electrical resistivity and magnetoresistance in the lanthanum hexaboride ($LaB_6$) doped by Ce and Ho in the regime of isolated magnetic impurity. Indeed, the low-temperature increase of electrical resistivity is the feature of the weak localization of charge carriers, whereas the negative magnetoresistance observed in $Ce_xLa_{1-x}B_6$ and $Ho_xLa_{1-x}B_6$ ($x=0.005-0.01$) at liquid helium temperatures corresponds to the formation of the spin-polaron type many-body states in $LaB_6$ near the magnetic rare-earth (RE) ions. Earlier, it was shown in [6] that the Kondo lattice model is also inapplicable for dense $CeB_6$ magnetic system. Indeed, at low temperatures, this commonly believed Kondo-lattice compound exhibits the weak localization of charge carriers ($\rho \sim T^{-0.37}$) and anomalous Hall effect [6], induced spin polarization [7], and ferromagnetic spin fluctuations [8, 9] rather than the Kondo screening of Ce ions' magnetic moment. Similar situation occurs in $CeAl_3$ and $CeCu_6$ intermetallic compounds with the record high values of the effective mass of heavy fermions ($m^*=1000 \div 1600\ m_0$, see [10,11]). With the decrease of temperature, these compounds which were usually treated as archetypal examples of the Kondo-lattice exhibit an additional magnetic response (ferromagnetic fluctuations) instead of the Kondo screening of the magnetic moment of $Ce^{3+}$ ions by itinerant electrons in the vicinity of the characteristic spin fluctuation temperature $T_K \sim 5$ K [12]. In addition, an *activation-type behavior of the Hall coefficient* was observed in these paramagnetic metals [12,13] (see Fig. 1). In [14], it was reported that the metal-insulator transition observed in the so called *Kondo insulators* $Tm_{1-x}Yb_xB_{12}$, but in the strong pulsed magnetic field up to 50T it is accompanied with arising of a component of magnetization corresponding to the many-body in-gap resonance in the electron density of states near the Fermi level, which also does not agree with the predictions of the Kondo model.

The aforementioned examples as well as the other ones reported in literature clearly demonstrate that at least in the SCES under discussion (*i*) the mechanism underlying the formation of heavy fermions in the vicinity of RE-ions is different from the Kondo one, (*ii*) the localization radius for these many-body states (heavy fermions) turns out to be nearly equal to the lattice constant (~5 Å) being much smaller than that of the "Kondo cloud" (>20 Å), and (3) the strong local spin fluctuations near rare earth ions lead to the formation of both heavy fermions and the spin-polarized nanodomains (ferrons, following the terminology of [15, 16]).

Currently, for the theoretical description of many-body effects in strongly correlated electron systems, the anisotropic Kondo model (AKM) is often used instead of the conventional Kondo approach. Application of various kind AKM approaches (see, e.g. [17-22]) and of the spin-boson Hamiltonian [23, 24] has quite recently provided an opportunity to develop an adequate description of the continuous quantum phase transition (for example, in [17, 21] the Bose-Fermi-Kondo model for an array of magnetic impurities coupled both to the fermionic reservoir and to the dissipative boson subsystem was used). It also allowed obtaining the values of the anomalous exponent characterizing the behavior of dynamic magnetic susceptibility, which are in a good agreement with the experimental data [17,21]. Moreover, the values of other parameters were found which are relevant to the non-Fermi-liquid behavior of the system at the quantum critical point (QCP) [18, 21]. In contrast to that, the results obtained by the calculations in the framework of the periodic Anderson model predict the scenario of the first order phase transition at QCP, which does not correspond to any actual situation [25, 26]. At the same time, the nature of the local (non-Kondo) spin fluctuations is not addressed at all in these models. In [17-22], for example, the fluctuations are described in terms of the coupling to the dissipative boson reservoir.

Taking into account that the anisotropic Kondo model is a special case of the spin-boson Hamiltonian describing the properties of the dissipative two-state system [23], it is natural to argue that the origin of spin fluctuations is related to the quantum oscillations of a heavy particle (magnetic ion) between the states in the double-well potential. The equivalence of the spin-boson Hamiltonian (see the diagram in inset of Fig. 1) and the AKM Hamiltonian (see Eqs. (3.58)-(3.65) in [23]) allows one to find the correspondence between the parameters of both these models.

It is worth noting that in $LaB_6$ and $CeB_6$ hexaborides, we have a loosely bound state of a rare-earth ion in the rigid covalent boron sub-lattice. The existence of boron vacancies leads to the displacements of $R^{3+}$ ions from the central positions within the large size crystallographic cavities formed by the $B_{24}$ cuboctahedrons (see Fig. 2). Eventually, this gives rise to the emergence of the double-well potentials. Such arising two-level systems are reliably manifest of themselves in $LaB_6$ and $Ce_xLa_{1-x}B_6$ at the measurements of the low temperature specific heat [27-28]. In contrast to that, the formation of the double-well potential in $CeAl_3$, $CeCu_6$, and other cerium-based intermetallic compounds seems to be related to the existence of several electronic configurations closely spaced in energy (quantum chemistry nature of the double-well potential). We believe that the activation type dependence of the Hall coefficient in the archetypal heavy-fermion metals $CeAl_3$ and $CeCu_6$ (Fig.1, [12]) is related just to the temperature-dependent tunneling of $Ce^{3+}$ ions between the levels in the neighboring potential wells. It is worth noting also that the quantum oscillations of $Ce^{3+}(Ho^{3+})$ ions in the double-well potential appeared in cavities of $B_{24}$ of the $LaB_6$ matrix (see inset of Fig. 1) allow us to interpret not only the formation of heavy fermions, but also the effects of weak localization and strong negative magnetoresistance reported in [5] for these both Kramers and non-Kramers RE-ions.

Let us also emphasize that the novel concept put forward here, which clarifies the mechanisms underlying the formation of the many-body states interprets in the natural way a strong scattering of charge carriers by spin fluctuations as well as arising of both antiferromagnetic and ferromagnetic local spin fluctuations, the nature of which is determined by the transitions between the quantum states in the double-well potential. It is clear that in the dense magnetic systems exhibiting quantum oscillations of rare-earth ions, one could expect the effect of spin polarization of the states in the $5d$ band accompanied by arising of nanosize ferromagnetic domains, which should take part in the formation of the complicated magnetic structure together with localized magnetic moments ($4f$ and $5d$ components of antiferromagnetic ordering).

In conclusion, we can say that in the analysis of applicability of different approaches to the description of strongly correlated electron systems, it is important to distinguish the Kondo effect as its manifesting itself in the formation of the singlet (nonmagnetic) ground state resulting from the spin-flip scattering of itinerant electrons by the localized magnetic moment of a rare-earth ion, and the widely used term "Kondo model", which is, in fact, equivalent to the *s-d* exchange model proposed for the first time by Vonsovskii [29]. At the same time, it was shown, e.g. in [5, 6], that the Kondo effect is irrelevant to the formation of many-body states of different nature, whereas the anisotropic *s-d* exchange model (or AKM) turns out to be suitable for the description of the many-body effects only so far as it is equivalent to the spin-boson Hamiltonian. However, if we do not want to lose the physical meaning of the obtained results, which can be hidden by the mathematical formalism, it is quite important to realize the difference mentioned above. In our opinion, it is not a good idea, to use the Kondo-related terminology since it completely changes the physical meaning of the phenomena under study and misleads experimentalists, discovering the Kondo resonances in the compounds with Kramers and non-Kramers magnetic 4$f$ ions (for $CeB_6$ and $PrB_6$, see e.g. [30,31]).

We thank K.I. Kugel' for collaboration and useful discussion. The work was supported by the program "Strongly Correlated Electron Systems" of the Division of Physical Sciences of the Russian Academy of Sciences and RFBR Project № 15-02-02553.

**References**

1. P.W. Anderson, **124**, 41 (1961)

2. P. Anderson, Rev. Mod. Phys. **50**, 191 (1978)

3. J.R. Schrieffer, P.A. Wolff, Phys. Rev. **149**, 491 (1966)

4. J. Kondo, Prog. Theor. Phys. **32**, 37 (1964)

5. N. E. Sluchanko, M. A. Anisimov, A. V. Bogach, V. V. Voronov, S. Yu. Gavrilkin, V. V. Glushkov, S. V. Demishev, V. N. Krasnorusskii, V. B. Filippov, N. Yu. Shitsevalova, JETP Letters **101**, 36 (2015).

6. N. E. Sluchanko, A. V. Bogach, V. V. Glushkov, S. V. Demishev, V. Yu. Ivanov, M. I. Ignatov, A. V. Kuznetsov, N. A. Samarin, A. V. Semeno, and N. Yu. Shitsevalova, JETP **104**, 120 (2007)

7. V. Plakhty, L.P. Regnault, A.V. Goltsev, S. Gavrilov, F. Yakhou, J. Flouquet, C. Vettier, S. Kunii, Phys. Rev. B, **71**, R11510 (2005)

8. S.V. Demishev, A.V. Semeno, A.V. Bogach, N.A. Samarin, T.V. Ishchenko, V.B. Filipov, N.Yu. Shitsevalova, N.E. Sluchanko, Phys. Rev. B, **80**, 245106 (2009).

9. Hoyoung Jang, G. Friemel, J. Ollivier, A.V. Dukhnenko, N.Yu. Shitsevalova, V.B. Filipov, B. Keimer, D.S. Inosov, Nature Mat. (2014) DOI: 10.1038/NMAT3976.

10. G. S. Stewart, Z. Fisk, M.S. Wire, Phys. Rev. B, 30, 482 (1984).

11. K. Andres, J.E. Graebner, H.R. Ott, Phys. Rev. Lett., 35**,** 1779 (1975).

12. D. N. Sluchanko, V. V. Glushkov, S. V. Demishev, G. S. Burhanov, O. D. Chistiakov, N. E. Sluchanko, J.Magn.Magn.Mat. **300,** 288 (2006)

13. N. E. Sluchanko, D. N. Sluchanko, N. A. Samarin, V. V. Glushkov, S. V. Demishev, A. V. Kuznetsov, G. S. Burhanov, O. D. Chistiakov, Low Temp. Phys. **35**, 544 (2009)

14. A. V. Bogach, N. E. Sluchanko, V. V. Glushkov, S. V. Demishev, A. N. Azarevich, V. B. Filippov, N. Yu. Shitsevalova, A. V. Levchenko, J. Vanacken, V. V. Moshchalkov, S. Gabani, and K. Flachbart, JETP **116**, 838 (2013).

15. E. L. Nagaev, JETP Lett. **6** (1), 18 (1967).

16. M. Yu. Kagan, K. I. Kugel, and D. I. Khomskii, JETP **93** (2), 470 (2001)

17. Jian-Xin Zhu, D.R. Grempel, Q. Si, Phys. Rev. Lett. **91**, 156404 (2003)

18. D.R. Grempel, Q. Si, Phys. Rev. Lett. **91**, 026401 (2003)

19. Lijun Zhu, S. Kirchner, Q. Si, A. Georges, Phys. Rev. Lett. **93**, 267201 (2004)

20. T. Senthil, M. Vojta, S. Sachdev, Phys. Rev. B **69**, 035111 (2004)

21. M. T. Glossop, K. Ingersent, Phys. Rev. Lett. **99**, 227203 (2007)

22. Ch. Thomas, A. S. da Rosa Simoes, C. Lacroix, J. R. Iglesias, B. Coqblin, arXiv:1404.6864v1 [cond-mat.str-el] 28 Apr 2014

23. A.J. Leggett, S. Chakravarty, A.T. Dorsey, M.P.A. Fisher, A. Garg, W. Zwerger, Rev. Mod. Phys. 59, 1 (1987)

24. T.A. Costi, Phys. Rev. Lett. **80**, 1038 (1998)

25. P. Sun, G. Kotliar, Phys. Rev. Lett. **91**, 037209 (2003)

26. P. Sun, G. Kotliar, Phys. Rev. B **71**, 245104 (2005)

27. M. Anisimov, A. Bogach, V. Glushkov, S. Demishev, N. Samarin, S. Gavrilkin, K. Mitsen, N. Shitsevalova, A. Levchenko, V. Filippov, S. Gabani, K. Flachbart, N. Sluchanko, Acta Physica Polonica A **126**, 350 (2014).

28. M. A. Anisimov, V. V. Glushkov, A. V. Bogach, S. V. Demishev, N. A. Samarin, S. Yu. Gavrilkin, K. V. Mitsen, N. Yu. Shitsevalova, A. V. Levchenko, V. B. Filippov, S. Gabani, K. Flachbart, and N. E. Sluchanko, JETP **116**, 760 (2013).

29. S. V. Vonsovskii, JETP **16**, 981 (1946)

30. S. Patil, G. Adhikary, G. Balakrishnan, K. Maiti, Solid St. Commun. **151**, 326 (2011)

31. S. Patil, G. Adhikary, G. Balakrishnan, K. Maiti, J. Phys.: Condens. Matter **23**, 495601 (2011)

**Figure captions**

**Fig. 1.** Temperature dependence of the Hall coefficient $R_H(T)$ for $CeAl_3$ and $CeCu_6$ compounds. We use the inverse logarithmic (activation-type) scale. Parameter $E_{a1}$ corresponds to the activation energy [12]. The inset illustrates the dependence of potential energy on the generalized coordinate $V(q)$ for the double-well potential (with the barrier height $V_0$). This dependence forms the basis for the spin-boson description of the dissipative systems characterized by two states (see [23]). In the diagram, we also show the heavy magnetic ion, performing quantum oscillations in the double-well potential.

**Fig. 2.** (a) unit cell of the $RB_6$, structure; (b) the change in the location of $R^{3+}$ ions near the boron vacancy (shown by large arrow). Small arrows indicate the displacements of rare-earth ions from the equilibrium positions in $B_{24}$ cuboctahedrons in direction from the boron vacancy, leading to the formation of the double-well potentials in four unit cells neighboring to the vacancy.

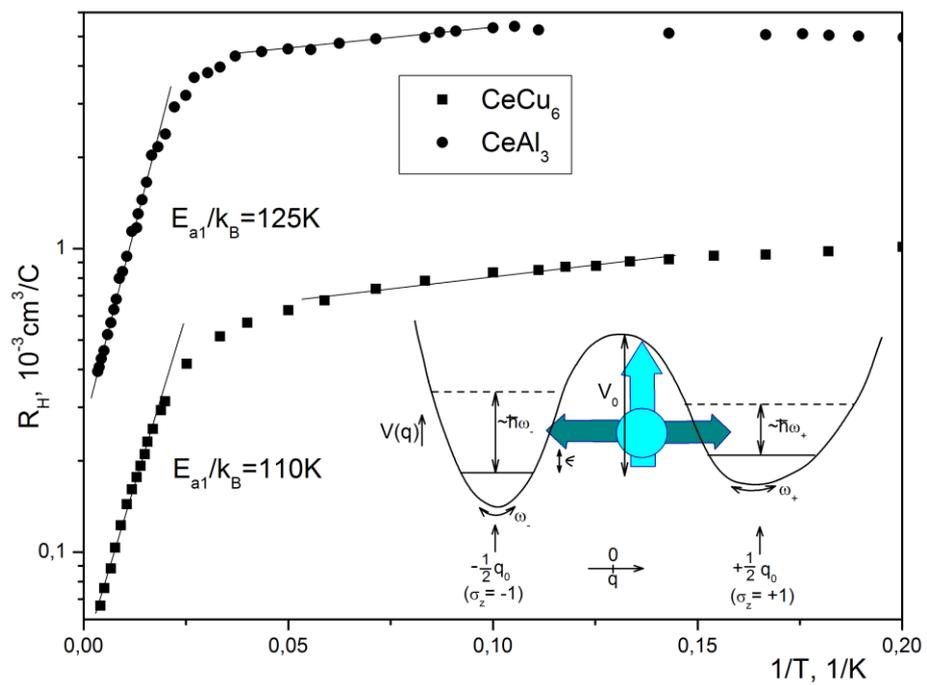

Fig.1.

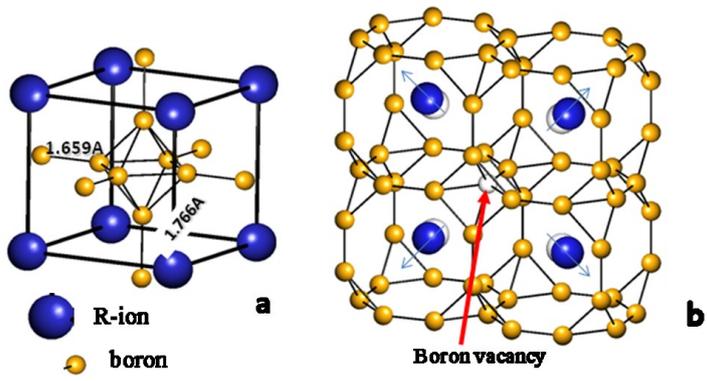

Fig. 2.